# Enhancing Safety in Automated Ports: A Virtual Reality Study of Pedestrian–Autonomous Vehicle Interactions under Time Pressure, Visual Constraints, and Vehicle Size

Yuan Che[1,2], Mun On Wong[3], Xiaowei Gao[4], Haoyang Liang[5], Yun Ye[1,2,6†]

[1] Faculty of Maritime and Transportation, Ningbo University, Ningbo, China

[2] Collaborative Innovation Center of Modern Urban Traffic Technologies, Southeast University, Nanjing, China

[3] Department of Civil and Environmental Engineering, University of Macau, Macao, China

[4] Department of Earth Science & Engineering, Imperial College London, London, UK

[5] College of Transportation Engineering and the Key Laboratory of Road and Traffic Engineering, Ministry of Education, Tongji University, Shanghai, China

[6] Centre for Transport Engineering and Modelling, Department of Civil and Environmental Engineering, Imperial College London, London, UK

† **Correspondence to**: yun.ye25@imperial.ac.uk

**Acknowledgement:** This research was supported by National Natural Science Foundation of China (Project No. 72501150), Zhejiang Provincial Natural Science Foundation of China (Grant No. LQN25E080011), Ningbo Natural Science Foundation (Grant No. 2024J440), and National "111" Centre on Safety and Intelligent Operation of Sea Bridges (Project No. D21013). The funders had no role in the study design, data collection and processing, manuscript preparation, or decision to publish.

**Authorship contribution statement**
**Yuan Che:** Conceptualization, Data curation, Formal analysis, Investigation, Methodology, Visualization, Writing-original draft. **Mun On Wong:** Visualization, Writing-review & editing. **Xiaowei Gao:** Validation, Writing-review & editing. **Haoyang Liang:** Writing-review & editing, Resources. **Yun Ye:** Conceptualization, Project administration, Funding acquisition, Supervision, Resources, Writing-review & editing.

**Declaration of competing interest**
The authors declare that they have no known competing financial interests or personal relationships that could have appeared to influence the work reported in this paper.

# Enhancing Safety in Automated Ports: A Virtual Reality Study of Pedestrian–Autonomous Vehicle Interactions under Time Pressure, Visual Constraints, and Varying Vehicle Size

**ABSTRACT**

Autonomous driving improves traffic efficiency but presents safety challenges in complex port environments. This study investigates how environmental factors, traffic factors, and pedestrian characteristics influence interaction safety between autonomous vehicles and pedestrians in ports. Using virtual reality (VR) simulations of typical port scenarios, 33 participants completed pedestrian crossing tasks under varying visibility, vehicle sizes, and time pressure conditions. Results indicate that low-visibility conditions, partial occlusions and larger vehicle sizes significantly increase perceived risk, prompting pedestrians to wait longer and accept larger gaps. Specifically, pedestrians tended to accept larger gaps and waited longer when interacting with large autonomous truck platoons, reflecting heightened caution due to their perceived threat. However, local obstructions also reduce post-encroachment time, compressing safety margins. Individual attributes such as age, gender, and driving experience further shape decision-making, while time pressure undermines compensatory behaviors and increases risk. Based on these findings, safety strategies are proposed, including installing wide-angle cameras at multiple viewpoints, enabling real-time vehicle–infrastructure communication, enhancing port lighting and signage, and strengthening pedestrian safety training. This study offers practical recommendations for improving the safety and deployment of vision-based autonomous systems in port settings.

# 1. Introduction

With the rapid advancement of global logistics and the accelerating development of smart ports, autonomous driving has emerged as a promising solution to improve operational efficiency and reduce costs in port logistics (Vaca-Recalde et al., 2024). Unlike the complex and dynamic nature of open-road environments, ports offer a relatively closed and structured setting, making autonomous driving comparatively easier to implement (Fiedler et al., 2019). With well-defined operational workflows, ports are considered high-value scenarios where autonomous driving is most likely to achieve early industrial deployment. Promoting the adoption of autonomous driving in ports can significantly reduce labor costs, improve operational efficiency, and enhance safety (Qin et al., 2020).

As ports transition toward fully unmanned operations, human workers and autonomous vehicles will inevitably continue to share operational space for the foreseeable future, where such conflicts remain prevalent (Qingdao West Coast New Area Government, 2022; Son et al., 2021). Recent deployments of driverless container trucks and yard tractors frequently occur in mixed-traffic scenarios, where machines and people operate in close proximity. For instance, in a 2022 pilot project at a Dutch terminal, autonomous electric yard tractors were deployed alongside other trucks, vehicles, and pedestrians during regular operations (Terberg Special Vehicles, 2022). In China, similar practices have been observed at major hubs such as Meishan and Mawan Ports, where unmanned vehicles coexist with ground personnel in stacking and transport areas (People's Daily Online, 2025; Häne et al., 2017; Cheng et al., 2021; Shit, 2020).

However, the integration of autonomous vehicles (AVs) into port operations also introduces unique safety requirements and challenges. First, port environments are often characterized by poor visibility due to both environmental and infrastructural factors (Lucio et al., 2024; Vaquero et al., 2018). Two distinct forms of visual limitation are particularly prevalent and pose different kinds of challenges: global (symmetric impairments) and local (asymmetric) obstructions. Global impairments refer to environmental conditions such as fog, heavy rain, and nighttime darkness, which uniformly reduce the visibility range for both AV sensors and pedestrians. These conditions diminish depth perception, blur motion cues, and lower the confidence of both pedestrians and AVs in evaluating spatial and temporal gaps during crossing (Wang et al., 2025). In contrast, local obstructions, such as stacked containers, parked trailers, cranes, or construction equipment introduce spatially asymmetric visual barriers that obstruct lines of sight in specific directions. These obstructions often lead to delayed detection and the sudden appearance of vehicles or pedestrians, significantly increasing the likelihood of conflict (Yang et al., 2016; Ge et al., 2020). In UK ports, for instance, over 30% of accidents are transport-related, often involving workers being struck in cargo yards due to limited real-time visibility around large objects (HSE, 2024; HSENI, 2020). While global impairments degrade overall situational awareness, local obstructions create blind zones that are sudden, uneven, and difficult to anticipate (Yoon et al., 2025; Macedo and Apolinario,2021). The coexistence of these two forms of visual constraints presents a compounded threat to safe pedestrian-vehicle interaction in port environments.

Moreover, the size of vehicles in ports may amplify safety risks. Autonomous container trucks and yard cranes are substantially larger and heavier than ordinary cars, meaning that any collision is likely to be far more severe (Amini et al., 2022). Statistical reviews have found that pedestrians are 40–50% more likely to be killed when struck by a heavy vehicle compared to a passenger car (Robinson et al., 2025). Beyond physical harm, vehicle size also influences human behavior and risk perception. Pedestrians tend to be more cautious and yield longer gaps when facing a large truck or platoon, reflecting a heightened perceived threat (Ye et al., 2024). In port settings, this dynamic can lead to risky decisions—for example, a worker might hesitate too long or dash abruptly if the looming presence of a massive autonomous vehicle triggers fear or urgency.

Port logistics operations function under stringent temporal demands, often driven by vessel schedules, cargo throughput quotas, and real-time vehicle dispatch systems. In such high-demand settings, time pressure becomes an omnipresent factor shaping both human and machine behavior (Human Element Industry Group, 2023). Pedestrians working in port environments may face implicit or explicit pressure to cross lanes quickly, particularly in operational scenarios that require repeated interactions with AVs. Time pressure is known to shorten decision times, reduce information scanning, and suppress cautious behavior (Hogenboom et al.,2021), factors that collectively undermine the quality of crossing decisions (Dhoke and Choudhary, 2025; Guo et al., 2024). Furthermore, AV systems themselves, optimized for efficiency, may prioritize schedule adherence unless explicitly designed with human-centered interaction protocols. The presence of time constraints thus interacts dangerously with the other two factors, impaired visibility and large vehicle presence forming a high-risk triad.

These challenges highlight the urgent need to systematically investigate pedestrian behavior and risk perception in automated port environments. Given the dynamic, visually constrained, and high-stakes nature of pedestrian–vehicle interactions in ports, conventional observational methods are often insufficient. To address this gap, the present study employs immersive virtual reality (VR) technology, a validated, safe, and repeatable approach for exploring human behavior in controlled yet realistic scenarios (Ye et al., 2020; Ye et al., 2023; Wong et al., 2023; Ye et al., 2024). This study proposes a VR-based experimental approach to investigate pedestrian–AV interaction behavior and risk in port environments. By simulating typical port traffic conditions, the VR platform allows for precise manipulation of key environmental factors, such as vehicle size, weather, lighting, visual obstructions, and time pressure. It enables the collection and analysis of multiple behavioral indicators, including gap acceptance, post-encroachment time (PET), waiting time, crossing time, and subjective risk perception, to reveal how pedestrians adapt their decision-making in response to environmental complexity and operational demands. The main contributions of this study are as follows:

◆ To the best of our knowledge, this study is among the first to systematically examine pedestrian-AV interaction behavior and risk in port environments, under effects of time pressure, visual constraints, and vehicle size.

◆ An immersive VR-based experimental approach is employed to simulate high-risk port conditions, enabling controlled, repeatable, and safe experimentation to quantify the effects of key contributory factors on pedestrian behavior.

◆ The study provides empirical evidence on pedestrian behavior in port settings, revealing how environmental and vehicle-related factors influence safety-critical decisions. These findings offer a scientific basis for optimizing AV perception algorithms, decision-making models, pedestrian–vehicle interaction design, and safety management strategies in port environments.

The remainder of the paper is structured as follows. Section 2 reviews related literature. Section 3 describes the methodology, including VR experiment, data collection, and data analysis. Section 4 presents the results. Section 5 discusses key findings and implications. Section 6 concludes the study and suggests directions for future research.

## 2. Literature review

### 2.1 Visual constraints and risk perception

Low visibility conditions markedly impair pedestrians' ability to accurately perceive and judge traffic, which in turn affects risk assessment. For instance, reduced lighting at night degrades visual processing and causes pedestrians to misjudge the speed of approaching vehicles (Balasubramanian and Bhardwaj, 2018). Similarly, heavy fog or haze can elevate perceived risk. Recent immersive simulations found that pedestrians report higher risk when crossing in foggy conditions (Kummeneje and Rundmo, 2019; Ferenchak and Abadi, 2021), especially at dusk (Zhu et al., 2025). These findings align with human factors theory on situational awareness: when critical visual cues are missing or unclear, pedestrians have lower perceptual confidence and may either hesitate in caution or make errors in judging safe gaps. Indeed, diminished visibility also eliminates or obscures non-verbal communication cues (e.g. eye contact or vehicle signals), further complicating the decision of when to cross in front of an approaching vehicle, an issue particularly relevant when the vehicle is an automated one lacking a human driver's cues. For instance, Hamka (2017) highlighted through fault tree analysis that shared zones between pedestrians and container vehicles are critical risk points in port terminals. Similarly, Zhu et al. (2025) found that in low-light conditions, pedestrians tend to perceive significantly greater danger when interacting with oncoming vehicles, which leads to more conservative gap acceptance behaviors.

A related constraint is physical occlusion in the environment, which can block pedestrians' and vehicles' line of sight. Prior studies on road crossings show that common occlusions (e.g. parked cars, buildings) significantly increase danger by hiding approaching traffic until the last moment (Morrongiello et al., 2018; Zhu et al., 2021). In port environments, occlusion problems are amplified: stacks of shipping containers, large cargo handling equipment, and crane structures create moving blind spots and obstructed sightlines far more extensive than typical urban settings. These port-specific obstacles can prevent pedestrians from seeing oncoming automated vehicles (and prevent the vehicles' sensors from detecting pedestrians) until very late. Despite general knowledge that occlusions undermine pedestrians' situational awareness and risk perception, virtually no research has examined this issue in a seaport or terminal context. While extensive research has clarified how visual limitations affect pedestrian safety in urban traffic, little is known about how such effects translate to port environments, where occlusions are often larger, dynamic, and spatially asymmetric posing unique perceptual and behavioral challenges yet to be empirically addressed.

**2.2 Vehicle size and behavioral response**

Empirical evidence confirms that pedestrians adjust their crossing behavior based on the size of the approaching vehicle. In traffic experiments, pedestrians tend to accept smaller gaps (shorter headways) when crossing in front of small vehicles, whereas they require larger gaps for buses or trucks (Petzoldt et al., 2017). In other words, a pedestrian might dart across with only a short time to spare when a compact car approaches but would wait for a much larger gap if the oncoming vehicle is a large truck. This behavior is reflected in differences in PET, a measure of the safety margin after crossing. Riskier, lower PET values tend to occur more frequently in the presence of smaller vehicles. From a perceptual standpoint, the well-known "size–arrival effect" likely contributes to this pattern: larger vehicles are perceived as arriving sooner (and appear more looming) than smaller ones (Petzoldt et al., 2017), prompting pedestrians to behave more cautiously (Yu et al., 2020). Additionally, larger vehicles evoke a greater perceived threat, a psychological deterrent that correlates with vehicle size. In contrast, smaller vehicles are intuitively judged as less intimidating, potentially leading to bolder crossing decisions (Petzoldt et al., 2017). These pedestrian–vehicle interaction dynamics—driven by vehicle size and perceived risk—are grounded in affordance theory and risk perception research. The looming presence of a large machine signals greater potential harm, prompting pedestrians to increase their safety buffer.

In the port environment, vehicle size and form factor vary dramatically and include some of the largest road-going machines. Specialized port vehicles such as straddle carriers, gantry cranes, and yard tractors are not only much larger than ordinary trucks or cars, but also operate in semi-structured environments with distinct movement patterns. One can expect that pedestrians (e.g., port workers on foot) may respond strongly to these imposing vehicles, possibly exhibiting highly conservative gap acceptance or avoidance behaviors. However, empirical research on pedestrian crossing behavior around such port vehicles—whether manually operated or autonomous—remains limited. While prior work suggests that vehicle size significantly shapes pedestrian caution and safety margins, most existing research fails to account for the oversized, industrial-scale vehicles operating in ports. This omission restricts our understanding of pedestrian–vehicle coordination in these high-risk environments.

**2.3 Time pressure and decision-making bias**

Time pressure is a well-documented factor that can bias pedestrians toward riskier decisions. Under time pressure—for example, when running late or trying to meet a deadline—pedestrians tend to reduce the time spent observing oncoming traffic and are more likely to accept smaller gaps between vehicles ([Morrongiello et al., 2015](#)). In virtual crossing experiments, participants under time constraints spent less time appraising traffic, selected more hazardous gaps, and crossed with minimal safety margins compared to when they were not pressed for time ([Morrongiello et al., 2015](#)).

Consistently, a recent systematic review identified "being in a hurry" as one of the strongest predictors of non-compliant road-crossing behavior ([Dhoke and Choudhary, 2023](#)). This suggests that urgency can override typical cautious behavior, potentially due to a shift in decision-making strategy: when under pressure, individuals rely more on fast heuristics and place greater momentary value on time savings than on safety—consistent with theoretical models of bounded rationality under stress. Notably, the influence of time pressure on pedestrian risk-taking appears consistent across age groups and settings ([Morrongiello et al., 2015](#)), suggesting a generalizable human factor rather than an isolated phenomenon. However, the literature in this area remains limited. Most existing studies have focused on simple mid-block crossings within controlled simulations, and few have directly examined how urgency affects finer aspects of pedestrian behavior, such as visual scanning patterns or gap estimation, in real-world contexts (Dhoke and Choudhary, 2023). While this limitation warrants caution in generalizing the findings, the prevailing trend is clear: urgency increases risk tolerance in pedestrian decision-making.

The relevance to port operations is direct: maritime and terminal environments often impose tight schedules. Such operational time pressure may lead port personnel on foot to take risks when crossing vehicle lanes or interacting with autonomous equipment, particularly if detours or delays could disrupt workflow (Human Element Industry Group, 2023). Under high time pressure, pedestrians in ports may also skip standard safety checks. For instance, they might cross without carefully verifying that an approaching autonomous vehicle has detected their presence, due to a cognitive bias toward quick action. However, no empirical studies have specifically examined how time pressure in port logistics scenarios affects pedestrian decision-making around AVs. Despite well-established behavioral patterns linking urgency to increased crossing risk, research in time-sensitive port environments remains limited. As a result, it is still unclear how operational deadlines may distort pedestrian judgment during interactions with autonomous systems.

Collectively, these research streams highlight the pressing need to investigate pedestrian behavior within the operational context of automated ports. While existing traffic safety literature offers relevant theoretical insights, it rarely accounts for the distinctive visual, vehicular, and temporal constraints present in container terminals and freight hubs. Addressing these overlooked conditions is essential for informing the development of AV technologies and safety strategies that reflect the real-world demands and pressures of port operations.

## 3. Methods

To recreate realistic port interaction scenarios for experimental study, this research first conducted a comprehensive investigation of actual port environments using publicly available sources, including official videos, photographs, and other visual materials. Based on these references, key spatial and operational features such as container stacks, autonomous trucks, road structures, and visibility constraints were extracted and reconstructed into a virtual port environment using Unreal Engine. This VR-based simulation platform enables controlled manipulation of multiple environmental factors while maintaining ecological validity, and is suitable for data collection in risk assessment.(Meng et al., 2025). Table 1 illustrates typical collision-risk scenarios observed in such contexts, for example, a yard tractor traveling at normal speed and approaching a worker who is about to cross the road (Son et al., 2021). These real-world observations and documented incidents indicate that pedestrian–vehicle overlap in port settings is neither rare nor negligible, but rather a systemic feature of current port operations. In light of this, it is essential to re-examine the safety implications of automated port environments, particularly through the lens of key risk factors affecting human–machine interactions.

**Table 1**. Illustrative human-vehicle interaction cases in port scenarios

| Port Scenario Snapshot | Visualization |
|---|---|
| A yard truck proceeds while a pedestrian approaches the same crossing point from an adjacent work area. The convergence zone between the pedestrian's path and the truck's trajectory is identified as an "interactive danger zone." | 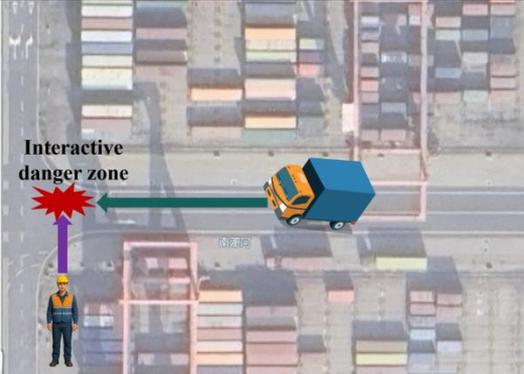 |

As illustrated in Fig. 1, the experimental design systematically varied five key factors commonly encountered in automated port operations: (1) weather conditions (sunny, rainy, foggy); (2) time of day (day vs. night); (3) presence of visual obstructions (e.g., stacked containers); (4) vehicle type (small vs. large autonomous truck platoons); and (5) time pressure (with or without crossing time limit). These factors were orthogonally combined to generate eight distinct simulation scenarios. Representative pedestrian crossing scenes were developed under obstructed conditions with both small and large vehicle platoons to examine how environmental stressors and vehicle characteristics influence pedestrian responses.

Participants were required to interact with the virtual port environment using head-mounted VR equipment and handheld controllers, navigating across traffic lanes while autonomous vehicle platoons approached. During each trial, two types of data were collected: (1) behavioral data, including gap acceptance (GA), post-encroachment time (PET), waiting time, and crossing duration; and (2) questionnaire data, covering risk perception, behavioral tendencies, immersion, and simulator sickness. These data were analyzed using one-way analysis of variance (ANOVA) to examine the main effects of the experimental conditions, and generalized linear mixed models (GLMMs) to account for repeated measures and individual variability. This combined analytical approach enabled a comprehensive assessment of how various environmental and individual factors influence pedestrian decision-making and safety outcomes in automated port settings.

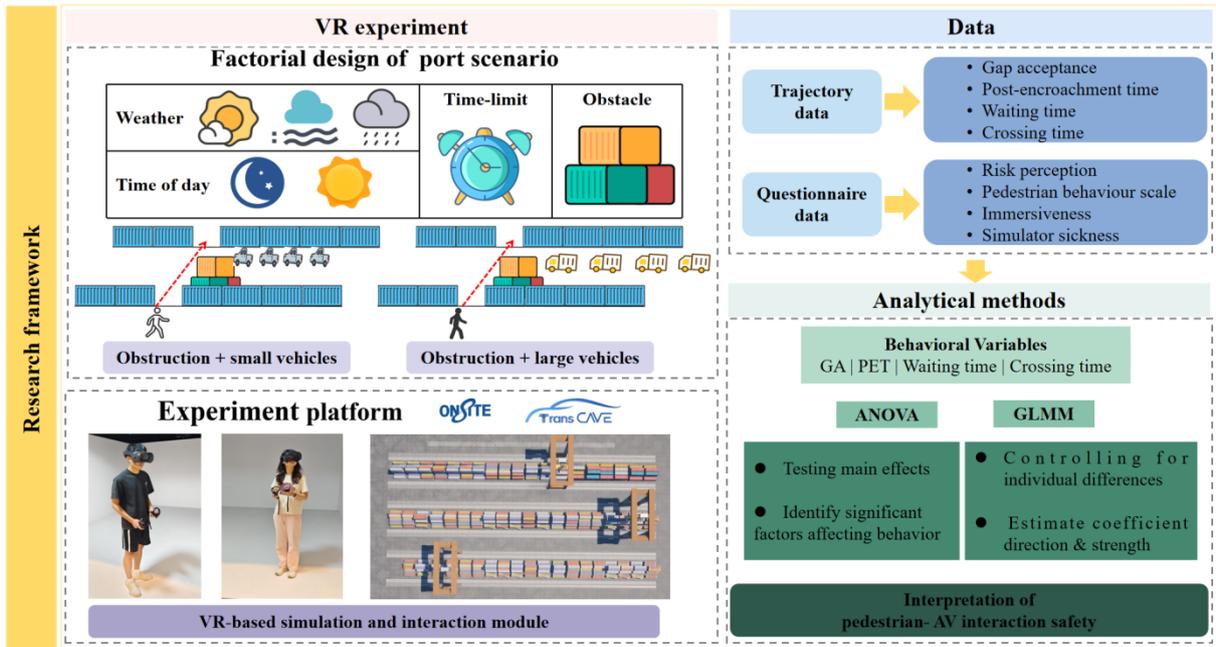

**Fig. 1.** Research framework.

### 3.1 Participants

A total of 33 participants (25 male, 8 female) were recruited from the university community to take part in the VR-based simulation experiment, with demographic information presented in Table 2. Although it was not feasible to recruit actual port workers due to access limitations, the sample was carefully selected to ensure contextual relevance. Notably, 72.7% of participants had academic or professional backgrounds in maritime and transportation, including students from maritime navigation, logistics, and transportation engineering programs. This background provided them with a foundational understanding of port operations, vehicle types, and spatial constraints, enabling them to make reasonably informed judgments in the simulated port environment.

Participants ranged in age from 20 to 28, with the majority (84.8%) being postgraduate students. Most held a valid driver's license (90.0%), and over 60% had at least one year of driving experience. These characteristics ensured that participants were familiar with vehicle behavior and traffic decision-making processes, thereby enhancing the ecological validity of their responses in the virtual pedestrian-vehicle interaction tasks.

**Table 2**. Demographic characteristics.

| Variable | Frequency | Proportion |
|---|---|---|
| **Gender** | | |
| Female | 8 | 24.2% |
| Male | 25 | 75.8% |
| **Age** | | |
| 20-22 | 9 | 27.3% |
| 23-25 | 18 | 54.5% |
| 26-28 | 6 | 18.2% |
| **Education level** | | |
| Undergraduate student | 5 | 15.2% |
| Postgraduate student or above | 28 | 84.8% |
| **Driver license** | | |
| Yes | 30 | 90.0% |
| No | 3 | 9.1% |
| **Years of driving experience** | | |
| 0 | 5 | 15.2% |
| <1 | 7 | 21.2% |
| 1-3 | 11 | 33.3% |
| 3-5 | 7 | 21.2% |
| >5 | 3 | 9.1% |
| **Professional experience in the field of maritime and transportation** | | |
| Yes | 24 | 72.7% |
| No | 9 | 27.3% |
| **Collision experience** | | |
| Yes | 10 | 30.3% |
| No | 23 | 69.7% |

## 3.2 Apparatus

The experimental platform utilizes Tongji University's Cave Automatic Virtual Environment (CAVE) laboratory (https://tops.tongji.edu.cn/jxpy/sypt.htm). The platform supports head-mounted VR devices. It is particularly well-suited for developing pedestrian-focused traffic experiments, allowing for the collection of behavioral and motion data across a range of scenarios such as urban streets, terminal interiors, and enclosed operational areas. To support this study, the research team constructed a virtual port environment using Unreal Engine, incorporating detailed visual elements including gantry cranes, containers, tower cranes, autonomous vehicles, and road networks, as shown in the simulation interface depicted in Fig. 2(a).

The pedestrian simulator in this experiment employs a stereoscopic head-mounted display (HMD), specifically the VIVE Focus 3. This device provides a resolution of 4896 × 2448 pixels (2448 × 2448 per eye) with a 90 Hz refresh rate, offering an immersive 360° VR experience. Participants navigated the virtual port using handheld controllers that allowed for free movement and orientation, as illustrated in Fig. 2(b). In addition, a workstation was used to initiate and terminate trials, present a third-person view of the virtual environment to participants, and manage data collection. The workstation setup includes an Intel Core i7-14700K CPU and an NVIDIA RTX 4070 Ti GPU with 12 GB of video memory.

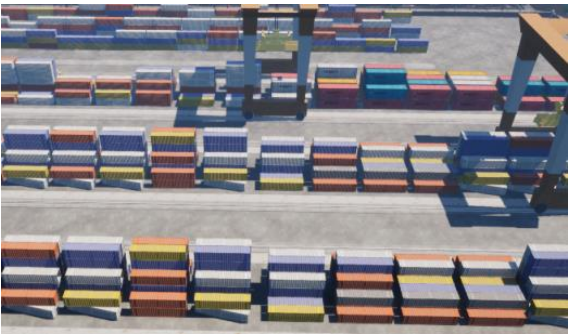 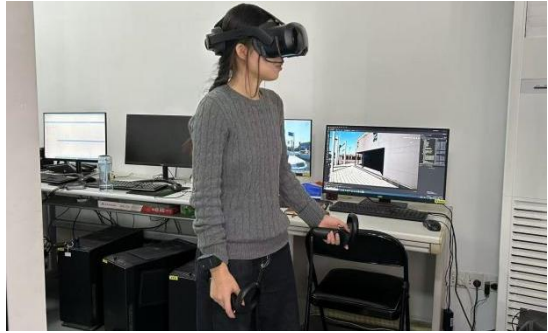

(a) Simulated port scenario     (b) Real-time monitoring workstation

**Fig. 2.** VR devices and scenes.

## 3.3 Experimental design

This study employed a design that systematically manipulated five key factors relevant to automated port environments: weather (sunny, rainy, foggy), time of day (day vs. night), visual obstruction (with or without stacked containers), time pressure (present vs. absent), and vehicle type (large vs. small autonomous truck platoons). These variables were orthogonally combined to generate eight representative scenarios, as shown in Table 3. This approach enabled an efficient and balanced investigation of how environmental and operational factors influence pedestrian crossing behavior and perceived safety.

**Table 3.** VR experiment scenario design.

| Scenario | Time | Weather | Obstacle | Time pressure | Vehicle size |
|---|---|---|---|---|---|
| 1 | Daytime | Sunny | Yes | No | Large |
| 2 | Daytime | Rainy | No | Yes | Large |
| 3 | Daytime | Foggy | Yes | Yes | Small |
| 4 | Daytime | Sunny | No | No | Small |
| 5 | Night | Sunny | Yes | Yes | Large |
| 6 | Night | Foggy | No | No | Large |
| 7 | Night | Rainy | Yes | No | Small |
| 8 | Night | Sunny | No | Yes | Small |

To ensure ecological validity, all scenarios were embedded within a realistic virtual port environment constructed in Unreal Engine, based on real-world port configurations (Son et al., 2021). Each crossing scene simulated a 10-meter-wide road flanked by stacked containers. Participants were required to cross from one side of the road to the other through this gap while interacting with an approaching AV platoon, as illustrated in Fig. 3.

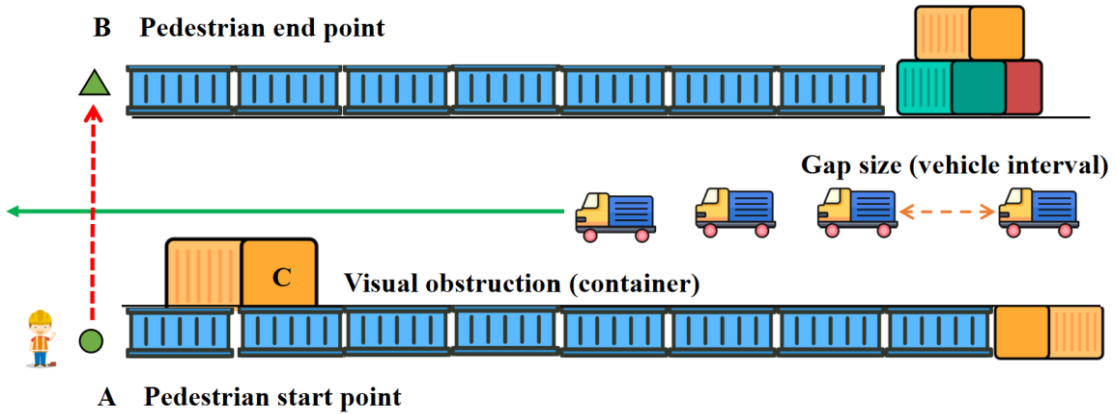

**Fig. 3.** Pedestrian crossing scenario.

Vehicle platoons consisted of 11 identical vehicles traveling at a constant speed, creating 10 inter-vehicle gaps. The vehicle dimensions differed between large (9.5 m × 4 m × 4.1 m) and small (5 m × 2 m × 2 m) platoons. To simulate realistic traffic flow, gap sizes were designed to gradually increase, with several shorter gaps intentionally inserted at specific positions (gaps 3, 5, 7, and 8) to examine gap selection behavior under varying time pressures (Paschalidis et al., 2018), as shown in Table 4.

Table 4. Gap size settings.

| Gap ID | 1 | 2 | 3 | 4 | 5 | 6 | 7 | 8 | 9 | 10 |
|---|---|---|---|---|---|---|---|---|---|---|
| Gap size (s) | 3 | 4 | 3.5 | 5 | 4.5 | 6 | 5.5 | 7 | 6.5 | 8 |

Time pressure was introduced by imposing a 40-second crossing deadline, which was determined through pilot testing. In time-limited scenarios, a countdown timer was displayed within the virtual environment and reinforced by auditory alerts every 10 seconds, simulating the sense of task urgency commonly experienced by port workers during high-paced operations.

To standardize the interaction, participants were instructed to wait until the first vehicle of the platoon had passed the line segment AB before initiating their crossing. This requirement ensured genuine engagement with the inter-vehicle gaps, the sizes of which were recorded as each participant's accepted gap, as illustrated in Fig. 3.

Each participant completed all eight scenarios in a randomized order. Prior to the formal trials, participants were given time to freely explore the virtual port environment to familiarize themselves with the navigation controls and spatial layout. This exploration phase ensured that participants—particularly those with maritime and transportation backgrounds—could reasonably simulate the decision-making processes encountered in actual port operations.

**3.3 Visibility-based deceleration trigger distance calibration**

To facilitate this investigation, a parameter termed the deceleration trigger distance was introduced, defined as the critical forward distance at which a vision-based autonomous vehicles, upon detecting a pedestrian, initiates braking to a complete stop. This parameter reflects the system's perception-reaction threshold and is directly influenced by the visibility conditions of the environment. Given that the perception range of vision-based autonomous vehicles is highly sensitive to lighting, weather, and occlusions, it was assumed that the deceleration trigger distance would vary across different scenarios. By applying fixed thresholds calibrated to specific visibility conditions, the study focuses on the behavioral dynamics of pedestrian-vehicle interactions, particularly pedestrians' crossing decisions and psychological responses without the confounding effects of adaptive sensor performance.

To determine this threshold empirically, a legibility calibration task was embedded within the VR environment. A standardized test chart containing printed text was placed along the autonomous vehicle's approach path. Under varying visibility conditions, including different weather scenarios (sunny, rainy, foggy) and times of day (day vs. night), the distance between the chart and the participant's viewpoint was gradually adjusted. For each condition, the maximum distance at which participants (pre-screened to exclude vision impairments) could clearly read the text was recorded. This reading distance was then used as a surrogate for the perceptual threshold at which mutual detection between pedestrian and vehicle occurs, and where a crossing or braking decision would be initiated.

This method draws on established practices in simulation-based human factors research, where legibility distance, the farthest point at which text or signs can be clearly read is widely used as a proxy for visual detection thresholds under varying visibility conditions. Prior studies have employed similar approaches to calibrate visibility and perception in driving simulators (Ting et al., 2008; Zhong et al., 2011), and to evaluate recognition distance of signage under degraded conditions in VR (Krösl et al., 2018). By analogy, using the maximum readable distance as the deceleration trigger point provides a perceptually grounded and empirically justifiable threshold for modeling worst-case pedestrian-vehicle interactions in low-visibility port environments.

By programming the autonomous vehicles to begin decelerating precisely at this threshold, the protocol achieves two key objectives:

(1) It ensures that braking is triggered only after the pedestrian would have been perceptually visible to a real-world vision-based autonomous vehicles, thereby avoiding premature or artificial reactions;

(2) It replicates worst-case interaction scenarios under limited visibility, thereby maximizing the impact of visibility conditions on pedestrian-vehicle interaction safety.

The calibrated deceleration trigger distances for each visibility condition are summarized in Table 5, and the calibration procedure is illustrated in Fig. 4.

**Table 5.** Deceleration trigger distance under different visibility conditions.

| Day of day | Weather | Deceleration trigger distance (m) |
|---|---|---|
| Daytime | Sunny | 9 |
| Daytime | Rainy | 8.5 |
| Daytime | Foggy | 7 |
| Night | Sunny | 8 |
| Night | Rainy | 6 |
| Night | Foggy | 5 |

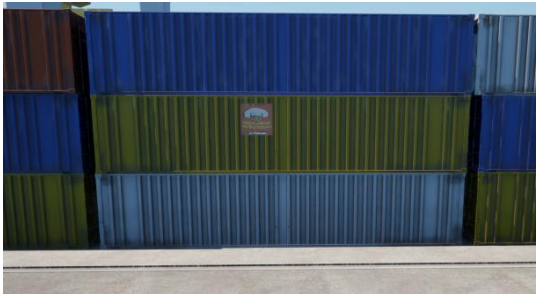
(a) Sunny, daytime condition

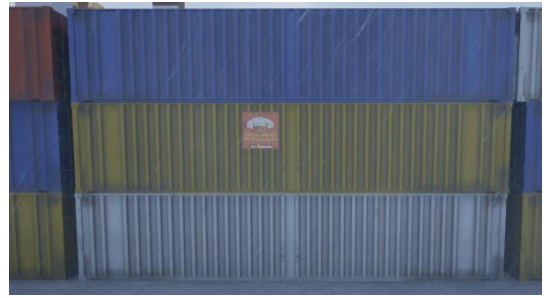
(b) Rainy, daytime condition

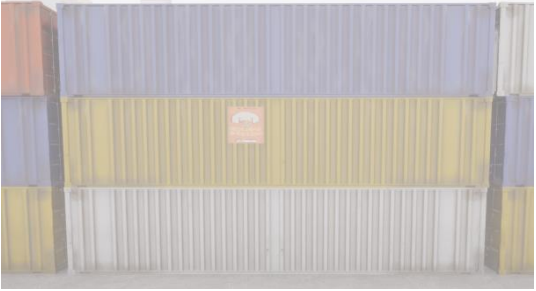
(c) Foggy, daytime condition

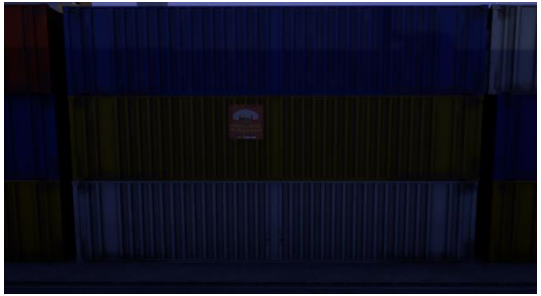
(d) Sunny, night condition

**Fig. 4.** Calibration of deceleration trigger distance.

The distances above refer to the autonomous vehicle's longitudinal direction of travel. However, as pedestrians may approach from the lateral side, it is also necessary to consider lateral visibility and response timing. Suppose the pedestrian moves laterally at a speed $v_{ped}$, and the autonomous vehicle travels forward at a speed $v_{veh}$. To ensure the autonomous vehicle reacts in time, the longitudinal distance it travels while the pedestrian moves into the decision zone must not exceed the zone's effective radius. Mathematically, if $t$ is the time it takes for the pedestrian to reach the autonomous vehicle's path (see Fig. 5), then the following inequality must be satisfied:

$$v_{veh}\ t \leq R \qquad (1)$$

where $R$ is the radial extent of the vehicle's perceptual zone. Expanding this relationship based on geometric assumptions leads to the following constraint:

$$v_{\text{veh}} \, \frac{X - \frac{1}{2}w}{v_{\text{ped}}} \leq R \qquad (2)$$

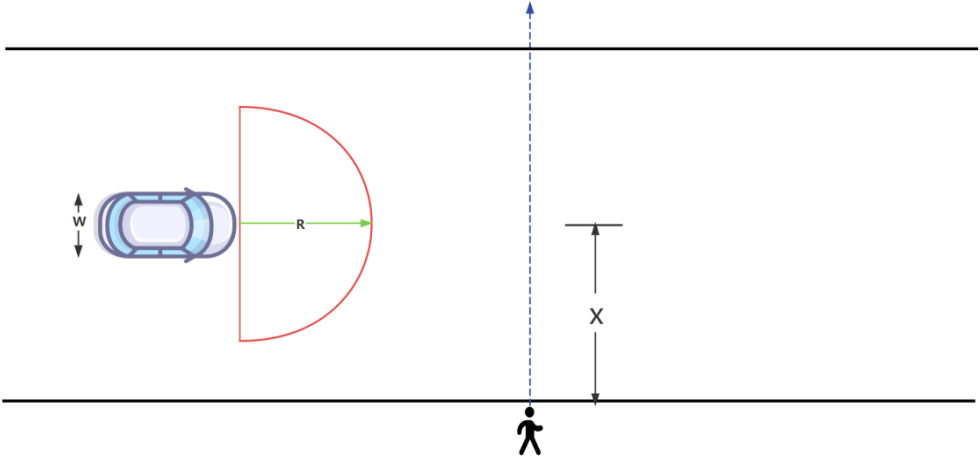

**Fig. 5.** Deceleration trigger distance.

This constraint ensures that the autonomous vehicle will initiate braking before the pedestrian reaches the critical collision zone, even during lateral crossings.

Furthermore, in certain scenarios, stacked containers were deliberately placed to occlude the vehicle's side view. Under such partial occlusion, the lateral perceptual range is effectively reduced. As illustrated in Fig. 6, the lateral trigger zone is assumed to shrink to half the vehicle's width, simulating realistic blind spots commonly encountered in port environments.

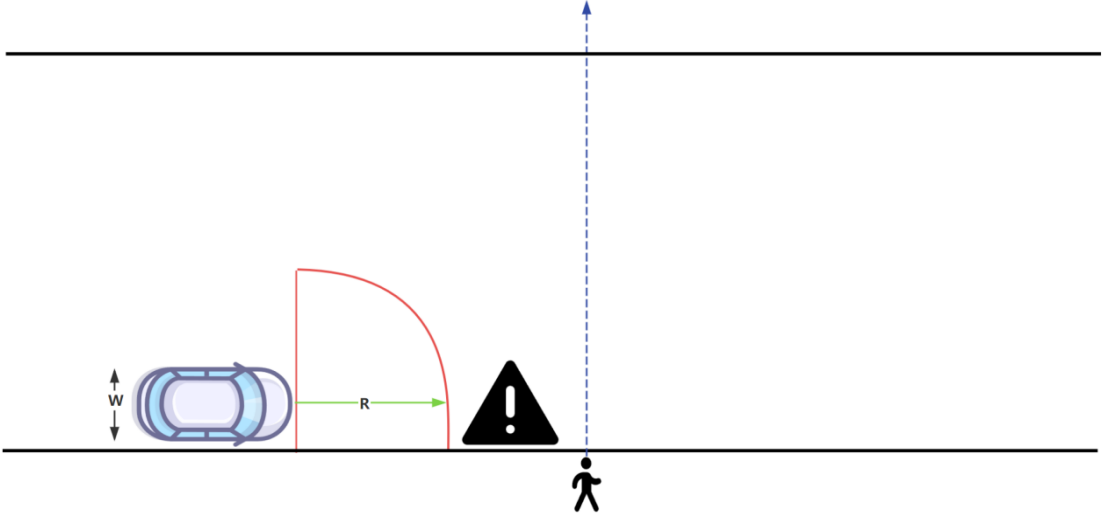

**Fig. 6.** Deceleration trigger distance under lateral visual obstruction.

### 3.4 Experimental procedure

The study was reviewed and approved by the Science and Technology Ethics Committee of Tongji University, and the experiment was conducted in the CAVE laboratory at Tongji University. A total of 33 participants completed the VR-based port interaction trials. Prior to the formal experiment, each participant underwent a health screening. Individuals who reported symptoms of discomfort or susceptibility to motion sickness were excluded. Participants then received a detailed briefing on the study's objectives, procedures, and safety instructions.

Before the formal trials began, all participants were allowed to freely explore the virtual port environment. This familiarization phase was intended to help participants develop spatial orientation and build confidence in completing the tasks. Once familiarized, participants completed the Virtual Reality Sickness Questionnaire (VRSQ; See Appendix A), a validated instrument for assessing motion sickness symptoms in virtual environments, with a focus on oculomotor discomfort and disorientation (Kim et al., 2018). Only those who reported no adverse symptoms proceeded to the main experiment.

Each participant completed all eight scenarios in a randomized order, as defined by the orthogonal experimental design (see Table 3). At the start of each trial, the participant's avatar was positioned at the designated point A, with a stacked container placed at point C in the obstruction condition (see Fig. 3). Participants were instructed to initiate crossing only after the first vehicle had passed line AB, ensuring genuine interaction with the inter-vehicle gaps.

Participants navigated the VR scene using handheld controllers, while the system continuously recorded the 3D coordinates and movement directions of both pedestrians and vehicles. In trials involving time pressure, a 40-second countdown timer and auditory alerts were provided to simulate urgency.

Upon completing each trial, participants were asked to provide a subjective risk perception (RP) assessment of the scenario, including perceived danger, likelihood of a collision, expected severity in the event of an accident, and concerns about autonomous vehicles (CAV; See Appendix B). Each item was rated on a 10-point Likert scale (1 = not at all, 10 = extremely). After completing all trials, participants filled out additional questionnaires, including the Multimodal Presence Scale (MPS; See Appendix C), a validated instrument for measuring physical, social, and self-presence in virtual environments (Makransky et al., 2017), and the Pedestrian Behavior Scale (PBS; See Appendix B) (Granié et al., 2013), which assessed behavioral tendencies and levels of immersion. The VRSQ was also re-administered at the end of the experiment to evaluate any delayed symptoms of simulator sickness.

Participants retained the right to withdraw from the experiment at any time if they experienced discomfort. The overall experimental workflow is illustrated in Fig. 7.

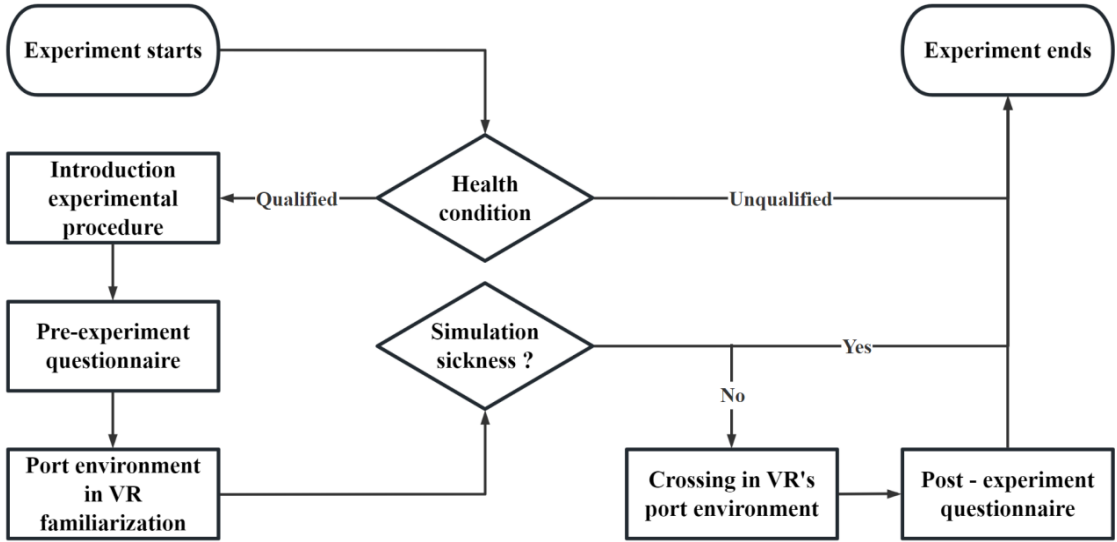

**Fig. 7.** Overall experimental workflow.

### 3.5 Dependent variables

To evaluate pedestrian safety and decision-making under different port conditions, this study defined four key behavioral outcome variables based on behavioral data and survey responses. In addition, demographic information (e.g., gender, age) was recorded for inclusion in mixed-effects modeling.

- **Gap acceptance (GA)** refers to a pedestrian's decision to cross an intersection or roadway segment after evaluating the temporal gap between vehicles. As an indicator of how pedestrians perceive available traffic gaps, GA serves as a critical metric for assessing both crossing behavior and perceived risk. Each participant's accepted gap size was therefore recorded.
- **Post-encroachment time (PET)** quantifies the temporal safety margin between a pedestrian and an oncoming vehicle after the pedestrian has entered the conflict zone. Specifically, when a pedestrian passes through the conflict point before the vehicle arrives, PET represents the remaining time buffer. This metric has been validated as a reliable surrogate for collision risk and provides methodological support for conflict-based safety assessments in traffic. Lower PET values indicate more severe conflict and a higher risk of collision, whereas higher values suggest safer interactions (Hermans et al., 2009).

- **Waiting time** is defined as the duration between a pedestrian's initial intent to cross and the actual moment they step onto the roadway. This variable reflects both risk assessment and decision-making latency.
- **Crossing time** refers to the interval between a pedestrian's first step onto the roadway and their arrival at the opposite curb. It captures both movement efficiency and the psychological or situational pressure experienced under different environmental conditions.

Taken together, these four variables provide a comprehensive behavioral profile of pedestrian-autonomous vehicle interactions in automated port environments.

### 3.6 Analytical methods

In this study, one-way ANOVA was employed to evaluate whether each experimental factor, including weather, time of day, obstacle presence, time pressure, and vehicle type, had a significant effect on pedestrian behavior metrics such as GA, PET, waiting time, and crossing time. This analysis served as an initial screening to identify which factors independently influence pedestrian behavior, thereby informing subsequent, more detailed analyses.

The GLMM extends the generalized linear model by incorporating both fixed effects and random effects, making it suitable for analyzing hierarchical or correlated data, as well as non-normally distributed outcomes. In this study, the GLMM was used to simultaneously estimate the average impact of experimental manipulations (fixed effects) and to account for individual variability among participants (random effects), providing a robust framework for analyzing pedestrian behavior under varied port-environment scenarios. The generalized linear mixed model can be written in the following form:

$$g(E(y_{ij})) = \beta_0 + \beta_1 X_{1ij} + \cdots + \beta_p X_{pij} + b_i \tag{3}$$

where $y_{ij}$ denotes the j-th observation in the i-th group, $E(y_{ij})$ represents the expected value of the dependent variable, and is linked to a linear combination of predictors via a specified link function $g(\bullet)$. The terms $\beta_0$, $\beta_1$, ..., $\beta_p$ are the fixed-effect coefficients, and $X_{1ij}$, $X_{pij}$ are the covariates associated with those fixed effects. The term $b_i$ represents the random effect component.

Specifically, the dependent variables include GA, PET, waiting time, and crossing time. These were modeled as functions of five fixed-effect factors: weather (clear, rain, fog), time of day (daytime or nighttime), obstacle presence (with or without containers), time pressure (present or absent), and vehicle type (large or small). Participant identity was included as a random effect to capture inter-individual variability in behavioral responses. This model structure allowed for the nested nature of the data, with repeated measurements taken across eight scenarios for each participant.

The GLMM was particularly appropriate for this study for two main reasons. First, it accommodates dependent variables that deviate from normality, such as GA and PET, by using flexible link functions. Second, it effectively models the within-subject correlations inherent in repeated-measures designs, thereby enabling valid statistical inferences.

## 4. Results

### 4.1 ANOVA analysis

To identify which experimental and demographic variables had significant effects on pedestrian behavior, one-way ANOVA was conducted on four key dependent variables: GA, PET, waiting time, and crossing time. This analysis served as a preliminary screening step to determine which factors should be retained for further modeling using GLMM. The complete statistical results are presented in Table 6, and the significant effects are visually illustrated in Fig. 8.

**Table 6.** One-way ANOVA results for the effects of experimental and demographic factors on pedestrian behavior.

| Independent variable | Dependent variable | $F$ | $\eta_p^2$ |
|---|---|---|---|
| Night | GA | 9.963** | 0.037 |
| Rainy | | 6.797* | 0.025 |
| Foggy | | 3.912* | 0.015 |
| Time pressure | | 31.322*** | 0.107 |
| Vehicle size | | 23.955*** | 0.084 |
| Age | | 2.89** | 0.083 |
| Transport practitioner | | 4.277* | 0.016 |
| Foggy | PET | 8.351** | 0.031 |
| Obstacle | | 5.645* | 0.021 |
| Age | | 2.323* | 0.068 |
| Years of driving experience | | 4.961* | 0.071 |
| Obstacle | | 11.784* | 0.043 |
| Time pressure | | 4.676* | 0.018 |
| Vehicle | Waiting time | 4.79* | 0.018 |
| Gender | | 10.397** | 0.038 |
| Age | | 4.076*** | 0.113 |
| Years of driving experience | | 3.184* | 0.047 |
| Collision experience | | 10.896* | 0.040 |
| Foggy | | 4.128* | 0.016 |
| Gender | Crossing time | 3.312* | 0.012 |
| Years of driving experience | | 2.433* | 0.036 |
| Night | | 119.084*** | 0.312 |
| Foggy | | 39.815*** | 0.132 |
| Obstacle | RP1 | 38.704*** | 0.129 |
| Time pressure | | 4.876* | 0.018 |
| Vehicle size | | 56.824*** | 0.178 |
| Night | | 108.963*** | 0.294 |
| Foggy | RP2 | 39.46*** | 0.131 |
| Obstacle | | 26.192*** | 0.091 |
| Vehicle size | | 44.049*** | 0.144 |
| Night | | 14.827*** | 0.054 |
| Rainy | | 17.361*** | 0.062 |
| Foggy | RP3 | 12.499*** | 0.046 |
| Obstacle | | 21.661*** | 0.076 |
| Vehicle size | | 72.804*** | 0.217 |
| Gender | | 6.448* | 0.024 |

$***: p<0.001; **: p<0.01; *: p<0.05$

Note: RP1, RP2, RP3: Risk perception 1, 2, 3 (See Appendix B).

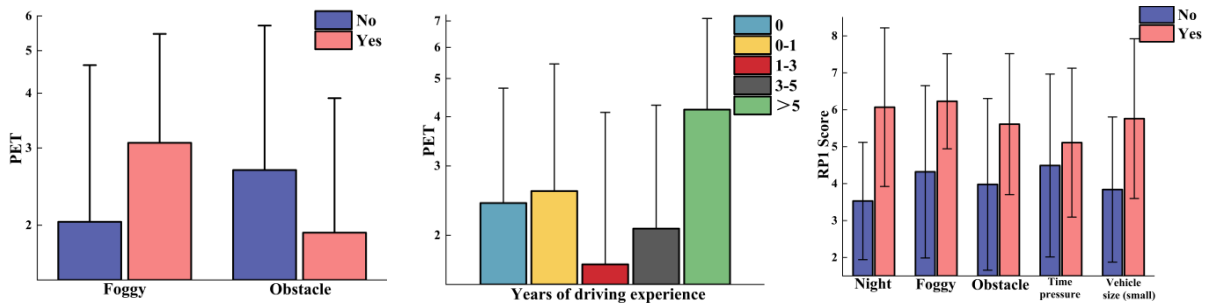

(a) PET  (b) RP1

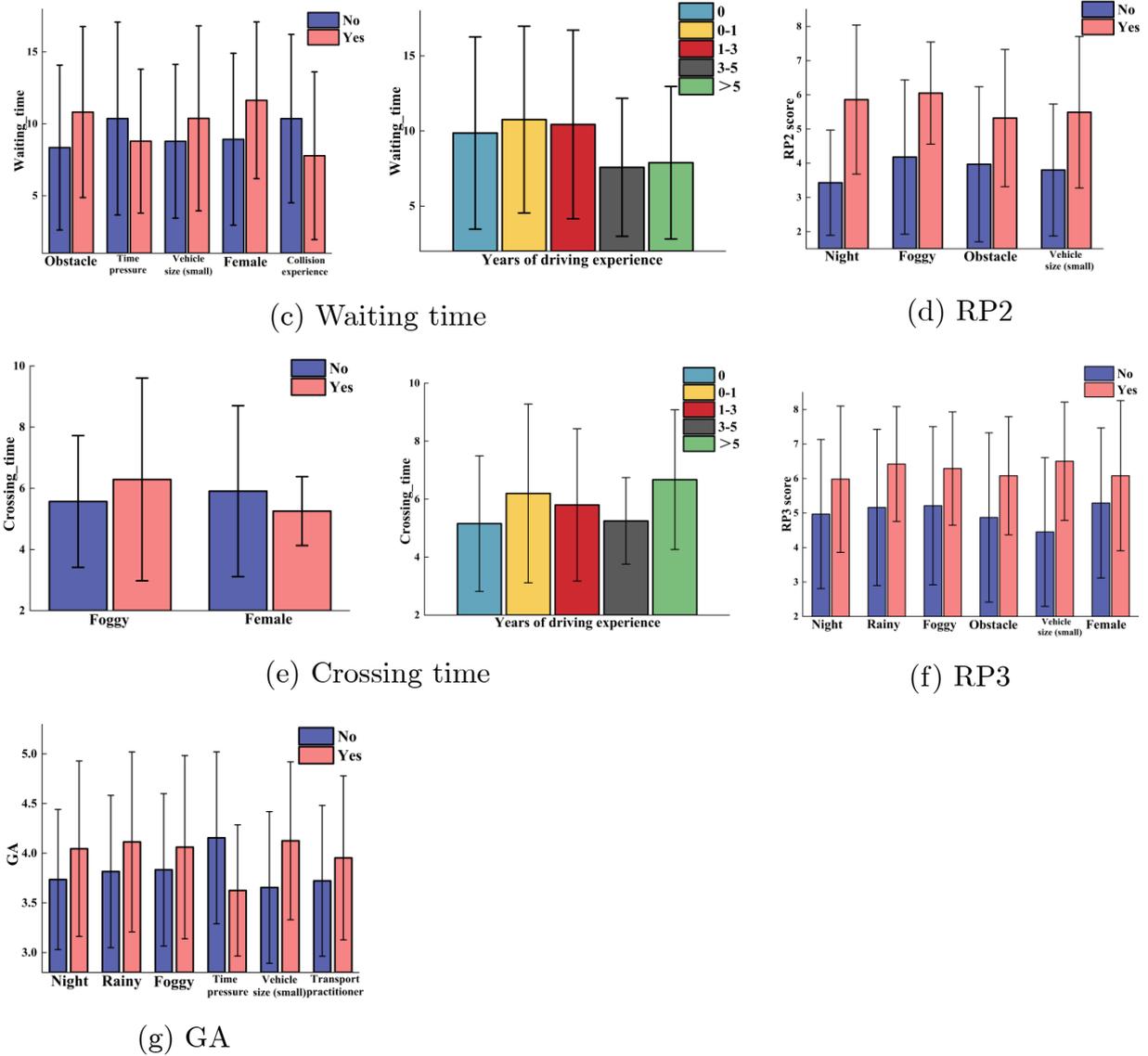

(c) Waiting time

(d) RP2

(e) Crossing time

(f) RP3

(g) GA

**Fig. 8.** Group comparisons of behavioral measures (GA, PET, waiting time, crossing time; risk perception (RP)) under significant experimental and demographic factors.

As shown in Fig. 8(g) and Table 6, gap acceptance was significantly influenced by both environmental and individual-level factors. Among all variables, time pressure had the strongest effect (F = 31.322, $\eta_p^2$ = 0.107), indicating that participants accepted significantly smaller gaps when a crossing deadline was imposed. This suggests that urgency can override risk considerations, leading to more aggressive crossing behavior. Vehicle size also had a substantial effect (F = 23.955, $\eta_p^2$ = 0.084): participants consistently accepted larger gaps when interacting with large autonomous trucks, reflecting elevated caution in response to the perceived threat posed by larger vehicles. In terms of visibility, rain (F = 6.797, $\eta_p^2$ = 0.037), fog (F = 3.912, $\eta_p^2$ = 0.015), and nighttime (F = 9.963, $\eta_p^2$ = 0.037) conditions all significantly increased accepted gap size, implying that reduced visual certainty prompts pedestrians to allow more buffer time before crossing. On the demographic side, age (F = 2.890, $\eta_p^2$ = 0.083) and transport practitioner (F = 4.277, $\eta_p^2$ = 0.016) were associated with more conservative gap selection, likely due to greater risk awareness or traffic knowledge.

As illustrated in Fig. 8(a) and Table 6, PET was most strongly influenced by foggy conditions (F = 8.351, $\eta_p^2$ = 0.031) and the presence of visual obstacles (F = 5.645, $\eta_p^2$ = 0.021). Demographically, age (F = 2.323, $\eta_p^2$ = 0.068) and years of driving experience (F = 4.961, $\eta_p^2$ = 0.071) were positively associated with PET, indicating that older and more experienced participants tended to maintain larger post-conflict safety margins.

A notable finding is that fog and visual obstruction had opposite effects on PET (Fig. 8(a)). While foggy conditions increased PET, container-induced occlusion significantly reduced it. This contrast can be attributed to the differing nature of visual impairment. Fog produces a global and symmetric reduction in visibility, prompting both pedestrians and vehicles to act more cautiously and allowing for longer observation and crossing buffers. In contrast, local obstructions (e.g., stacked containers) result in asymmetric information loss: pedestrians cannot see oncoming vehicles until they appear suddenly, and pure-vision autonomous vehicles may likewise fail to detect pedestrians approaching from behind the obstruction. These limitations reduce reaction time and lead to more immediate pedestrian-vehicle conflicts, thereby decreasing PET. This finding suggests that localized occlusions pose a greater safety risk than general low visibility, emphasizing the importance of incorporating redundant perception mechanisms to preserve safety margins in occluded environments.

As seen in Fig. 8(c) and Table 6, visual obstruction emerged as the strongest predictor of waiting time before crossing (F = 11.784, $\eta_p^2$ = 0.043). Participants took longer to initiate crossing when their view was blocked by containers or other obstacles, suggesting elevated uncertainty. Vehicle size (F = 4.790, $\eta_p^2$ = 0.018) and time pressure (F = 4.676, $\eta_p^2$ = 0.018) also had significant effects. Larger vehicles prompted hesitation, while the presence of a countdown timer led to shorter decision latency.

Among individual variables, age (F = 4.076, $\eta_p^2$ = 0.113) showed a strong inverse relationship: older participants waited less, which may appear counterintuitive but could reflect faster situational assessment. In contrast, collision experience (F = 10.896, $\eta_p^2$ = 0.04) and driving experience (F = 3.184, $\eta_p^2$ = 0.047) were associated with longer waiting times, possibly due to heightened risk sensitivity.

Crossing time was relatively stable across environmental conditions, with only fog showing a marginally significant effect (F = 4.128, $\eta_p^2$ = 0.016), as depicted in Fig. 8(e) and Table 6. However, gender (F = 3.312, $\eta_p^2$ = 0.012) and driving experience (F = 2.433, $\eta_p^2$ = 0.036) did exhibit small but significant effects. On average, female participants and those with less driving experience crossed more quickly, which may reflect defensive movement strategies: spending less time exposed in the conflict zone.

Overall, the ANOVA results indicate that time pressure, vehicle size, and visibility-related factors (fog, night, and visual obstruction) consistently influence pedestrian crossing decisions and perceived safety margins. Age, driving experience, and prior collision history also play important roles in behavioral variability. These findings justify the inclusion of all five experimental variables as fixed effects in the subsequent GLMM analysis. To account for repeated measures and behavioral differences across individuals, participant-specific random intercepts were included in the model. This approach allowed the GLMM to control for individual-level variability while focusing on estimating the average effects of environmental conditions on pedestrian behavior.

### 4.2 Generalized linear mixed model
#### 4.2.1 Gap acceptance
The GLMM results, as summarized in Table 7, show that several environmental variables significantly influenced pedestrians' accepted gap sizes. Time pressure had the strongest negative effect ($\beta$ = -0.135, $p$ < 0.001), indicating that pedestrians under time constraints tended to accept smaller gaps, likely prioritizing task urgency over safety. Vehicle size also had a strong positive effect ($\beta = 0.114$, $p < 0.001$), suggesting that larger vehicles elicited greater caution, consistent with intuitive threat perception.

Table 7. GLMM results for pedestrian gap acceptance behavior.

| Parameters | Coefficient | Std. err. | z-statistics | p-value | 95% conf. interval |
|---|---|---|---|---|---|
| **Fixed effects** | | | | | |
| Night | 0.063 | 0.030 | 2.12 | 0.034 | [0.005, 0.122] |
| Rainy | 0.119 | 0.027 | 4.42 | 0.000 | [0.066, 0.171] |
| Foggy | 0.091 | 0.033 | 2.8 | 0.005 | [0.027, 0.155] |
| Obstacle | 0.025 | 0.024 | 1.03 | 0.301 | [-0.022, 0.071] |
| Time pressure | -0.135 | 0.018 | -7.7 | 0.000 | [-0.170, -0.101] |
| Vehicle size | 0.114 | 0.028 | 4.05 | 0.000 | [0.059, 0.169] |
| Female | 0.012 | 0.054 | 0.22 | 0.825 | [-0.094, 0.118] |
| Age | -0.001 | 0.014 | -0.07 | 0.948 | [-0.028, 0.026] |
| Driving license | -0.034 | 0.085 | -0.4 | 0.686 | [-0.201, 0.132] |
| Years of driving experience | 0.011 | 0.024 | 0.48 | 0.629 | [-0.034, 0.058] |
| Education level | -0.048 | 0.074 | -0.64 | 0.519 | [-0.192, 0.097] |
| Professional experience | 0.063 | 0.051 | 1.24 | 0.216 | [-0.037, 0.163] |
| Collision experience | 0.016 | 0.044 | 0.36 | 0.716 | [-0.071, 0.103] |
| Violation | -0.001 | 0.014 | -0.06 | 0.953 | [-0.029, 0.027] |
| Error | -0.022 | 0.018 | -1.2 | 0.232 | [-0.058, 0.014] |
| Lapse | -0.005 | 0.006 | -0.89 | 0.372 | [-0.016, 0.006] |
| Aggressive | 0.011 | 0.012 | 0.9 | 0.370 | [-0.013, 0.035] |
| Positive | -0.027 | 0.012 | -2.29 | 0.022 | [-0.050, -0.004] |
| RP1 | 0.026 | 0.014 | 1.79 | 0.073 | [-0.002, 0.053] |
| RP2 | -0.017 | 0.012 | -1.47 | 0.141 | [-0.040, 0.006] |
| RP3 | -0.013 | 0.009 | -1.45 | 0.147 | [-0.031, 0.005] |
| CAV | 0.015 | 0.009 | 1.68 | 0.092 | [-0.002, 0.033] |
| **Random effects** | | | | | |
| var(_cons) | 0.006 | 0.002 | | | [0.003, 0.011] |

Note: Violation, error, lapse, aggressive and positive are items in PBS and CAV refers to participants' concerns about autonomous vehicles (See Appendix B).

Visibility-related factors also shaped gap acceptance. Gap sizes increased significantly under rainy ($\beta = 0.119$, $p < 0.001$), foggy ($\beta = 0.091$, $p <$ 0.01), and nighttime ($\beta = 0.063$, $p < 0.05$) conditions, reflecting pedestrians' compensatory behavior when perceptual uncertainty was high.

Among individual factors, positive behavioral tendencies were associated with significantly smaller accepted gaps ($\beta = -0.027$, $p = 0.022$), possibly reflecting faster decision-making and greater confidence during crossing. This pattern suggests that individuals who typically demonstrate courteous, cooperative, and attentive pedestrian behaviors may also be more confident in assessing traffic gaps and initiating movement. Such participants likely possess stronger situational awareness and action readiness, enabling them to make quicker decisions and navigate crossings with reduced hesitation, even under varying environmental conditions. Other personal attributes, such as age, gender, and driving experience, showed no significant influence on GA in the model.

The random intercept variance for participants was estimated at 0.006, indicating modest but non-negligible individual differences in baseline gap acceptance behavior. This supports the inclusion of participant-specific random effects in the model, accounting for within-subject correlations and enhancing the precision of fixed-effect estimates.

In summary, GA was primarily shaped by external risk cues, notably time pressure, vehicle size, and visibility, while internal traits like behavioral disposition played a secondary but meaningful role. These findings emphasize the importance of designing pedestrian-autonomous vehicle interaction systems that can adapt to task urgency and perceptual ambiguity in real-world environments.

**4.2.2 Post-encroachment time**

The GLMM results in Table 8 show that visual obstacles had a significant negative effect on post-encroachment time (PET) ($\beta = -0.492$, $p = 0.043$), indicating that container-induced occlusions reduced the safety buffer between pedestrians and vehicles at the conflict point. This suggests that localized obstructions compromise mutual visibility, leading to shorter reaction times and tighter pedestrian-vehicle interactions. By contrast, foggy conditions showed a positive but non-significant effect ($\beta = 0.371$, $p = 0.229$), implying that while fog may encourage caution, its impact on PET is less consistent and more context-dependent.

Table 8. GLMM results for post-encroachment time.

| Parameters | Coefficient | Std. err. | z-statistics | p-value | 95% conf. interval |
|---|---|---|---|---|---|
| **Fixed effects** | | | | | |
| Night | -0.027 | 0.285 | -0.1 | 0.924 | [-0.586,0.532] |
| Rainy | -0.121 | 0.273 | -0.44 | 0.657 | [-0.657,0.414] |
| Foggy | 0.371 | 0.308 | 1.2 | 0.229 | [-0.233,0.975] |
| Obstacle | -0.492 | 0.243 | -2.02 | 0.043 | [-0.969,-0.015] |
| Time pressure | -0.163 | 0.168 | -0.97 | 0.335 | [-0.493,0.168] |
| Vehicle size | -0.449 | 0.271 | -1.65 | 0.098 | [-0.981,0.083] |
| Female | 1.032 | 0.324 | 3.19 | 0.001 | [0.397,1.667] |
| Age | 0.164 | 0.081 | 2.04 | 0.042 | [0.006,0.322] |
| Driving license | -0.400 | 0.437 | -0.91 | 0.361 | [-1.257,0.458] |
| Years of driving experience | 0.315 | 0.133 | 2.37 | 0.018 | [0.055,0.576] |
| Education level | -0.256 | 0.425 | -0.6 | 0.546 | [-1.088,0.576] |
| Professional experience | -0.402 | 0.276 | -1.46 | 0.144 | [-0.942,0.138] |
| Collision experience | 0.035 | 0.229 | 0.15 | 0.880 | [-0.414,0.483] |
| Violation | -0.007 | 0.077 | -0.09 | 0.925 | [-0.158,0.144] |
| Error | -0.016 | 0.113 | -0.14 | 0.889 | [-0.237,0.205] |
| Lapse | 0.013 | 0.032 | 0.41 | 0.679 | [-0.05,0.077] |
| Aggressive | -0.015 | 0.073 | -0.2 | 0.839 | [-0.158,0.128] |
| Positive | -0.094 | 0.064 | -1.48 | 0.139 | [-0.219,0.031] |
| RP1 | 0.046 | 0.129 | 0.36 | 0.720 | [-0.207,0.300] |
| RP2 | -0.080 | 0.102 | -0.78 | 0.434 | [-0.280,0.120] |
| RP3 | 0.098 | 0.079 | 1.24 | 0.216 | [-0.057,0.252] |
| CAV | 0.003 | 0.081 | 0.04 | 0.970 | [-0.157,0.163] |
| **Random effects** | | | | | |
| var(_cons) | 0 | - | - | - | - |

Note: Violation, error, lapse, aggressive and positive are items in PBS and CAV refers to participants' concerns about autonomous vehicles (See Appendix B).

Vehicle size showed a marginally significant negative effect on PET ($\beta$ = -0.449, $p$ = 0.098), suggesting that pedestrians may cross closer in time to large vehicles, potentially due to delayed decision-making or longer traversal distances. Other environmental factors, including nighttime, rain, and time pressure, did not significantly affect PET in the model.

At the individual level, gender, age, and driving experience were all significant predictors. Female participants had significantly higher PET values ($\beta$ = 1.032, $p$ = 0.001), reflecting a more cautious crossing style. In addition, age ($\beta$ = 0.164, $p$ = 0.042) and years of driving experience ($\beta$ = 0.315, $p$ = 0.018) were both positively associated with PET, indicating that older and more experienced individuals tended to leave larger temporal buffers when crossing in front of vehicles.

In summary, PET was most strongly influenced by visual obstacles and individual caution-related traits, while global visibility conditions such as fog and rain played a lesser role. These findings highlight the elevated safety risks posed by asymmetric visibility and underscore the need for autonomous vehicle perception systems to account for blind spots and occluded pedestrians, particularly in container terminal environments.

### 4.2.3 Waiting time

As shown in Table 9, visual obstruction was the strongest predictor of waiting time before crossing ($\beta$ = 0.249, $p$ = 0.005). When containers blocked the line of sight, pedestrians exhibited longer hesitation before initiating movement, likely due to increased uncertainty about approaching vehicles. In contrast, time pressure significantly reduced waiting time ($\beta$ = -0.152, $p$ = 0.020), suggesting that urgency shortened decision latency, even when visual information was limited.

Table 9. GLMM results for waiting time.

| Parameters | Coefficient | Std. err. | z-statistics | p-value | 95% conf. interval |
|---|---|---|---|---|---|
| Fixed effects | | | | | |
| Night | 0.041 | 0.111 | 0.37 | 0.715 | [-0.177,0.259] |
| Rainy | 0.083 | 0.101 | 0.82 | 0.410 | [-0.115,0.281] |
| Foggy | 0.060 | 0.121 | 0.5 | 0.620 | [-0.177,0.296] |
| Obstacle | 0.249 | 0.089 | 2.8 | 0.005 | [0.075,0.424] |
| Time pressure | -0.152 | 0.066 | -2.32 | 0.020 | [-0.280,-0.023] |
| Vehicle size | 0.055 | 0.101 | 0.55 | 0.585 | [-0.143,0.254] |
| Female | 0.029 | 0.201 | 0.14 | 0.886 | [-0.366,0.423] |
| Age | -0.169 | 0.051 | -3.29 | 0.001 | [-0.27,-0.068] |
| Driving license | 0.286 | 0.315 | 0.91 | 0.364 | [-0.331,0.902] |
| Years of driving experience | 0.011 | 0.088 | 0.13 | 0.898 | [-0.161,0.184] |
| Education level | 0.579 | 0.273 | 2.12 | 0.034 | [0.044,1.114] |
| Professional experience | -0.021 | 0.189 | -0.11 | 0.910 | [-0.391,0.348] |
| Collision experience | -0.277 | 0.165 | -1.68 | 0.093 | [-0.601,0.046] |
| Violation | -0.003 | 0.053 | -0.06 | 0.951 | [-0.107,0.100] |
| Error | 0.046 | 0.069 | 0.67 | 0.505 | [-0.089,0.180] |
| Lapse | -0.025 | 0.021 | -1.18 | 0.239 | [-0.067,0.017] |
| Aggressive | 0.003 | 0.046 | 0.07 | 0.942 | [-0.087,0.094] |
| Positive | -0.060 | 0.044 | -1.38 | 0.168 | [-0.146,0.025] |
| RP1 | 0.043 | 0.054 | 0.8 | 0.424 | [-0.063,0.150] |
| RP2 | 0.003 | 0.047 | 0.06 | 0.951 | [-0.088,0.094] |
| RP3 | -0.014 | 0.033 | -0.44 | 0.662 | [-0.078,0.050] |
| CAV | 0.039 | 0.033 | 1.2 | 0.231 | [-0.025,0.103] |
| **Random effects** | | | | | |
| var(_cons) | 0.080 | 0.028 | | | [0.040, 0.158] |

Note: Violation, error, lapse, aggressive and positive are items in PBS and CAV refers to participants' concerns about autonomous vehicles (See Appendix B).

Age was a significant negative predictor of waiting time ($\beta$ = -0.169, $p$ = 0.001), indicating that older participants made quicker crossing decisions. This may reflect faster risk assessment or more experience navigating traffic. In contrast, education level was positively associated with waiting time ($\beta$ = 0.579, $p$ = 0.034), implying that individuals with higher education tended to act more cautiously. Collision history also showed a marginally significant negative effect ($\beta$ = -0.277, $p$ = 0.093), suggesting that those with prior accident experience may adopt more decisive strategies in ambiguous situations.

Age demonstrated opposite effects on waiting time and PET. As age increased, pedestrians tended to initiate crossing more quickly (shorter waiting time), but also maintained a larger temporal buffer after leaving the conflict zone (longer PET). This pattern suggests that older participants, while more decisive in initiating movement (possibly due to greater familiarity with traffic dynamics or reduced hesitation), also exhibited stronger risk-avoidance strategies during the crossing itself. Rather than delaying action, they appeared to prioritize efficient movement combined with conservative timing, ensuring that they cleared the vehicle's path well before its arrival. These findings reflect a behavioral style that is both experienced and precautionary, involving earlier decisions along with greater safety margins.

Other variables, including weather, vehicle size, and most personal traits, had no statistically significant effects on waiting time. This suggests that pedestrian hesitation is driven primarily by visual access to the environment, task urgency, and individual cognitive or experiential traits, rather than broader environmental context.

Overall, waiting time was jointly shaped by environmental visibility constraints and individual decision tendencies. Obstructed views triggered more hesitation, while time pressure and older age reduced deliberation time. These findings highlight the need for autonomous vehicle systems to interpret hesitation behavior as an indicator of uncertainty, particularly when pedestrian visibility is restricted.

### 4.2.4 Crossing time

In the GLMM analysis, as presented in Table 10, gender emerged as a significant predictor of crossing time. Female participants crossed more quickly than male participants ($\beta = -0.230$, $p = 0.016$). This may reflect a defensive movement strategy aimed at reducing exposure in potentially risky environments. Additionally, positive behavioral tendencies ($\beta = -0.051$, $p = 0.014$) and higher lapse scores ($\beta = -0.023$, $p = 0.022$) were associated with shorter crossing durations, indicating that individuals with either more impulsive or more proactive behavioral profiles tended to complete crossings more rapidly.

**Table 10.** GLMM results for crossing time.

| Parameters | Coefficient | Std. err. | z-statistics | p-value | 95% conf. interval |
|---|---|---|---|---|---|
| Fixed effects | | | | | |
| Night | -0.073 | 0.062 | -1.18 | 0.238 | [-0.193,0.048] |
| Rainy | -0.047 | 0.055 | -0.86 | 0.392 | [-0.154,0.060] |
| Foggy | 0.003 | 0.067 | 0.05 | 0.959 | [-0.129,0.136] |
| Obstacle | 0.005 | 0.049 | 0.11 | 0.915 | [-0.092,0.102] |
| Time pressure | -0.022 | 0.036 | -0.62 | 0.533 | [-0.093,0.048] |
| Vehicle size | -0.106 | 0.058 | -1.83 | 0.067 | [-0.220,0.007] |
| Female | -0.230 | 0.095 | -2.41 | 0.016 | [-0.417,-0.043] |
| Age | -0.036 | 0.025 | -1.46 | 0.145 | [-0.084,0.012] |
| Driving license | 0.093 | 0.149 | 0.62 | 0.535 | [-0.200,0.386] |
| Years of driving experience | 0.019 | 0.042 | 0.45 | 0.654 | [-0.063,0.100] |
| Education level | 0.121 | 0.130 | 0.93 | 0.351 | [-0.133,0.375] |
| Professional experience | 0.023 | 0.090 | 0.26 | 0.797 | [-0.154,0.200] |
| Collision experience | -0.102 | 0.078 | -1.31 | 0.192 | [-0.255,0.051] |
| Violation | 0.011 | 0.025 | 0.42 | 0.672 | [-0.039,0.060] |
| Error | -0.034 | 0.032 | -1.06 | 0.288 | [-0.098,0.029] |
| Lapse | -0.023 | 0.010 | -2.28 | 0.022 | [-0.043,-0.003] |
| Aggressive | 0.025 | 0.022 | 1.18 | 0.240 | [-0.017,0.068] |
| Positive | -0.051 | 0.021 | -2.46 | 0.014 | [-0.091,-0.010] |
| RP1 | 0.038 | 0.028 | 1.36 | 0.175 | [-0.017,0.094] |
| RP2 | -0.004 | 0.024 | -0.15 | 0.877 | [-0.05,0.0420] |
| RP3 | -0.004 | 0.018 | -0.24 | 0.809 | [-0.04,0.0310] |
| CAV | 0.043 | 0.018 | 2.34 | 0.019 | [0.007,0.0780] |
| **Random effects** | | | | | |
| var(_cons) | 0.015 | 0.006 | | | [0.007, 0.034] |

Note: Violation, error, lapse, aggressive and positive are items in PBS and CAV refers to participants' concerns about autonomous vehicles (See Appendix B).

Environmental variables such as time of day, weather, obstruction, and time pressure did not show significant effects on crossing time. This suggests that once pedestrians committed to crossing, environmental differences played a minimal role in their traversal speed. However, a marginal effect was observed for vehicle size ($\beta = -0.106$, $p = 0.067$), implying that larger vehicles may have subtly influenced crossing urgency.

The random intercept variance was estimated at 0.015 (95% CI [0.007, 0.034]), reflecting modest individual differences in baseline crossing speed. Incorporating participant-level random effects therefore remained appropriate for controlling intra-subject correlations and improving model fit.

Overall, crossing time was primarily influenced by individual behavioral traits rather than environmental conditions. This suggests that post-decision execution is more strongly governed by personal action styles than by external stimuli.

## 5.Discussion
### 5.1 Visual constraints: global vs. local effects

The experimental results demonstrate that visual constraints influence pedestrian–AV interactions in markedly different ways depending on whether the limitation is global or local in nature. In scenarios involving local visual obstruction, such as stacked containers positioned adjacent to the crossing path, PET decreased significantly. This reduction indicates a smaller temporal buffer between the pedestrian clearing the conflict point and the vehicle's arrival. Although participants exhibited longer pre-crossing hesitation under obstruction, the gain in waiting time did not compensate for the compressed PET. The likely explanation lies in the asymmetric nature of visibility loss: one party may remain completely undetected until the moment of entry into the conflict zone, leaving both sides with minimal reaction time.

In contrast, global visibility impairments (including fog and low-light conditions) were associated with more conservative gap acceptance. For example, gap acceptance increased under fog and rain, suggesting a deliberate extension of safety margins when the entire scene is uniformly degraded. This behavioral adjustment is consistent with findings reported by Zhu et al. (2025), yet it diverges from the results of Wang et al. (2025), who observed smaller gaps and higher exposure risk in urban nighttime conditions. The divergence likely reflects the structured and predictable spatial organization of port operations in our simulation, which may facilitate more measured decision-making under uniformly poor visibility compared with the dynamic, heterogeneous environment of public roads.

The distinction between global and local impairments has important mechanistic implications. Global impairments are symmetrical, affecting both pedestrian and vehicle perception in a predictable manner, thereby prompting mutual behavioral adaptation. Local obstructions, on the other hand, generate sudden and uneven visibility loss, which often prevents timely mutual detection. The asymmetric and abrupt nature of such occlusions explains why PET deteriorated despite longer decision latency.

The pronounced risk associated with local obstructions underscores the need for perception systems in port-based AVs that can compensate for asymmetric visibility. Effective strategies include the integration of multi-sensor fusion to detect occluded pedestrians, deployment of infrastructure-based surveillance at blind corners, and provision of wearable localization devices for ground personnel. These measures could mitigate the elevated conflict risk inherent to the spatial characteristics of container terminals.

**5.2 Vehicle size: risk perception and caution amplification**

The experimental findings demonstrate that vehicle size exerts a substantial influence on pedestrian crossing behavior in automated port settings. When interacting with large autonomous truck platoons, participants exhibited significantly more conservative decision-making compared with scenarios involving small vehicles. This was reflected in a greater mean accepted gap and a tendency to traverse the carriageway more quickly once the crossing was initiated. In addition, the one-way ANOVA indicated longer pre-movement hesitation in the presence of large vehicles. These results indicate a behavioral adjustment characterized by prolonged assessment before stepping into the conflict zone, followed by an accelerated crossing phase, suggesting an intentional strategy to minimize exposure time in proximity to large vehicles.

These behavioral tendencies are consistent with previous observational studies. Mohan and Chandra (2021) reported a marked reduction in gap acceptance rate in traffic streams dominated by heavy vehicles, relative to streams of smaller vehicles under identical gaps, while Tyndall (2021, 2023) found that collisions involving large vans or light trucks are associated with disproportionately severe outcomes for pedestrians. The consistency between our controlled VR findings and field-based evidence suggests that pedestrians' heightened caution is not merely perceptual but rather an adaptive response to objectively elevated risk. In port environments, this behavioral pattern is likely further amplified by the operational characteristics of large vehicles, including greater mass, higher front-end profiles, and limited maneuverability, particularly when operating in platoons.

Beyond perceived risk, the physical characteristics of large vehicles create substantive challenges for mutual detection. Simulation analyses by Jagtap and Jermakian (2025) show that large trucks have extensive blind zones, particularly during turning maneuvers, which impede both driver- and sensor-based recognition of pedestrians. In our experiment, such detection uncertainty may have led participants to shorten their crossing time once they had committed, thereby minimizing time spent on the vehicle's path. Psychological factors may also contribute, as suggested by the "safety anxiety" documented by Fabricius et al. (2022) in interactions with heavy trucks and the higher perceived danger ratings reported by Rankavat and Gupta (2023). These factors together provide a plausible explanation for the consistently greater caution observed in our large-vehicle scenarios.

The findings suggest that large AVs should be deployed with enhanced safety provisions, including expanded sensor coverage to mitigate blind zones and the integration of highly visible external human-machine interfaces (eHMI) to communicate yielding intentions. Operational measures, such as adjusting platoon routing, moderating speeds in pedestrian-accessible areas, and separating heavy vehicle flows from common pedestrian paths, can further reduce both perceived and actual risk. Such measures would not only address physical safety concerns but may also alleviate the heightened psychological stress experienced by pedestrians in the presence of large vehicles.

**5.3 Time pressure: urgency and risk trade-off**

The experimental results clearly demonstrate that time pressure alters pedestrian crossing behavior in ways that reduce safety margins. When a countdown was imposed, participants accepted significantly smaller temporal gaps and initiated crossings more quickly, as indicated by shorter waiting times. Although the model did not show a statistically significant increase in collisions or near-miss incidents, the reduction in PET under time pressure indicates that the temporal buffer between pedestrians and approaching vehicles was compressed. This suggests that urgency prompted participants to prioritize task completion over maintaining a generous safety margin.

These behavioral adjustments are consistent with the simulator-based findings of Tian et al. (2022), who reported that participants instructed to hurry selected smaller gaps, stepped into the roadway earlier, and spent less time in the conflict zone. Çinar et al. (2022) further observed that prolonged waiting can generate an "accumulated" form of time pressure, leading to similar risk-taking tendencies even in the absence of explicit deadlines. The synthesis by Dhoke and Choudhary (2023) reinforces the generality of this pattern, identifying "being in a hurry" as one of the most consistent predictors of pedestrian non-compliance. In our port-based simulation, these tendencies emerged despite the absence of external traffic complexity, indicating that the underlying mechanism was the cognitive effect of time constraint rather than environmental unpredictability.

The result also points to a likely shift in decision-making strategy under urgency. In scenarios with a visible countdown, the reduced waiting times and smaller accepted gaps suggest that participants may have allocated more attention to the temporal constraint itself than to assessing the trajectory and speed of the approaching vehicles. This interpretation is consistent with Colley et al. (2022), who reported that repeated exposure to countdown timers can condition pedestrians to initiate movement earlier, sometimes without adequate situational appraisal. Such reliance on temporal prompts may be particularly problematic in automated environments where vehicle-to-pedestrian communication (e.g., eHMI) could be misinterpreted or absent, as highlighted by Hochman et al. (2024).

In high-throughput port operations, where tight schedules are common, time pressure should be treated as a controllable safety variable. Strategies could include building modest schedule buffers to reduce the need for hurried crossings, avoiding excessive use of visual urgency prompts in mixed-traffic areas, and ensuring that AV systems communicate their yielding intentions in a clear and unambiguous manner. These measures would help prevent operational efficiency targets from inadvertently fostering riskier pedestrian behavior.

## 5.4 Implications

The behavioral patterns identified in this study have direct implications for improving pedestrian-AV safety in automated port environments. The effects of asymmetric visibility, vehicle size, and time pressure observed in the VR experiments underscore the need for a multi-pronged approach that integrates engineering design, operational enforcement, and targeted safety education.

**(1) Engineering**

The findings confirm that localized visual obstructions, such as stacked containers or parked equipment, can impair pedestrian–vehicle coordination more severely than global environmental factors such as fog. To address this risk, autonomous vehicles should be equipped with perception systems capable of extending detection beyond the direct line of sight. Multi-sensor fusion architectures combining radar, LiDAR, and camera inputs can enhance recognition accuracy in occluded zones, allowing AVs to respond promptly to pedestrians emerging from behind obstacles.

Complementary modifications to the physical environment can further strengthen situational awareness. Measures such as convex mirrors at blind corners, proximity warning systems, and visual alerts in mixed-traffic areas can help both AVs and pedestrians anticipate potential conflicts. Deploying real-time localization systems through networked wearable devices for all ground personnel would enable continuous position tracking, allowing AVs to adapt operations dynamically.

Wherever possible, physical segregation of pedestrian and vehicle flows remains the most effective safeguard. Dedicated walkways, fixed barriers, or grade-separated crossings can eliminate direct conflicts. In shared zones, visibility and predictability should be prioritized through clear crosswalk markings, adequate lighting, and dynamic signage to delineate safe crossing points and movement paths.

**(2) Enforcement**

The experimental results indicate that time pressure can lead to riskier crossing decisions, particularly during periods of intense operational demand. Adjustments to dispatch schedules and workflow sequencing that incorporate temporal buffers could reduce the need for hurried movements between operational areas.

Enforcing speed limits for all vehicles, whether autonomous or manually operated, in pedestrian-accessible zones is essential to reducing kinetic energy and increasing the time available for hazard detection and avoidance. Clearly defined right-of-way rules should be established, giving pedestrians priority at designated crossings and programming AVs to yield accordingly.

Operational protocols should also address visibility-related hazards. AVs should use both visual and auditory warnings when approaching blind corners or intersections, while pedestrians should avoid entering lanes with obstructed sightlines unless visibility is ensured. Mandating the use of connected localization devices for all workers would enhance both real-time collision avoidance and post-incident analysis.

Promoting a safety-first culture is equally important. Workers should be encouraged to report hazards without fear of reprisal, and supervisors should be empowered to temporarily suspend operations when unsafe conditions are identified. The adoption of "Stop Work Authority" policies, as seen in some leading ports, provides a model for embedding this principle in daily operations.

**(3) Education**

Education is fundamental to embedding safety awareness and ensuring the effective application of engineering and policy interventions. This study showed that pedestrian risk perception and crossing behavior vary systematically with vehicle size, visibility, and urgency. Training should therefore address decision-making strategies under these conditions, particularly in the presence of automated systems.

Instructional content should include the operational characteristics and limitations of AVs, especially their perception capabilities and potential blind zones under obstructed or low-visibility conditions. Safety briefings should emphasize consistent use of marked crossings, avoiding rushed decisions under time pressure, and waiting for clear and verifiable crossing signals.

Finally, training programs should explain the role and benefits of wearable localization devices, highlighting how their use enhances both individual safety and the overall situational awareness of the automated system. Clear understanding of these functions can improve worker compliance and integration of the technology into daily practice.

## 6. Conclusion

This study systematically examined pedestrian interactions with vision-based autonomous vehicles in complex port environments using immersive VR simulations. By incorporating key environmental variables such as weather, lighting, obstacle presence, vehicle size, and time pressure, along with individual pedestrian characteristics, the research revealed how these factors jointly influence risk perception and crossing behavior. The analyses demonstrated that low-visibility conditions and localized visual obstructions substantially affect pedestrian decision-making. These effects result in longer waiting time and larger accepted gaps, but also lead to reduced post-encroachment time, thereby narrowing the safety margin during pedestrian-vehicle interactions. Personal attributes including age, gender, and driving experience were also shown to influence behavioral outcomes, with time pressure further increasing the likelihood of risky decisions. Furthermore, vehicle size emerged as a crucial factor shaping pedestrian behavior. Larger vehicles elicited more conservative decisions, including larger accepted gaps and faster crossing speeds. These findings suggest that both visual systems and interaction strategies for large autonomous trucks must be calibrated to pedestrian risk sensitivity.

The findings highlight the limitations of relying solely on visual perception for autonomous vehicle navigation in ports and emphasize the importance of integrated safety strategies. Recommendations include the installation of wide-angle and elevated cameras to mitigate blind spots, the development of vehicle-to-infrastructure communication systems for real-time pedestrian tracking, and the improvement of port infrastructure such as lighting and signage. Furthermore, targeted safety training for port workers is essential to enhance situational awareness and reduce risks in human-machine interactions.

In summary, this study provides a comprehensive behavioral framework and practical design guidance for improving the safety and effectiveness of vision-based autonomous systems in port settings. By aligning technological solutions with environmental conditions, vehicle characteristics, and human cognitive responses, the study offers actionable pathways for advancing safe and efficient autonomous operations in high-risk real-world environments.

People's Daily Online. (2025, April 1). Smart and green dual-driven: Port construction enters the "fast lane" of development. *People's Daily Online*. Retrieved from: https://baijiahao.baidu.com/s?id=1828162052473225452&wfr=spider&for=pc. [2025-06-14].

Piyalungka, S., Kanitpong, K., & Karoonsoontawong, A. (2025). Pedestrian gap acceptance behavior at unsignalized mid-block crossing under mixed traffic conditions. *IATSS Research*, 49(2), 105-113.

Qin, K., Wang, B., Zhang, H., Ma, W., Yan, M., & Wang, X. (2020). *Research on application and testing of autonomous driving in ports* (No. 2020-01-5179). SAE Technical Paper.

Qingdao West Coast New Area Government. (2022, November 23). Investigation report on the "April 26" general vehicle injury accident of Qingdao Heshengjia Long Industrial and Trade Co., Ltd. in Huangdao District. Retrieved from: https://www.xihaian.gov.cn/zwgk/bmgk/qyjj/gkml/zdgz/dcbg/202212/t20221202_6542100.shtml.[2025-06-14].

Rankavat, S., & Gupta, V. (2023). Risk perceptions of pedestrians for traffic and road features. *International Journal of Injury Control and Safety Promotion*, 30(3), 410-418.

Rezwana, S., & Lownes, N. (2024). Interactions and behaviors of pedestrians with autonomous vehicles: A synthesis. *Future Transportation*, 4(3), 722-745.

Robinson, E., Edwards, P., Laverty, A., & Goodman, A. (2025). Do sports utility vehicles (SUVs) and light truck vehicles (LTVs) cause more severe injuries to pedestrians and cyclists than passenger cars in the case of a crash? A systematic review and meta-analysis. *Injury Prevention.*

Shanghai Daily. (2024, November 16). SAIC L4 intelligent container trucks "land" at Peru's Qiankai Port, assisting in the construction of South America's first smart port. *Shanghai Daily*. Retrieved from: https://baijiahao.baidu.com/s?id=1815847281557005784&wfr=spider&for=pc. [2025-06-14].

container terminals at Kaohsiung port. *Proceedings of the Institution of Mechanical Engineers, Part M: Journal of Engineering for the Maritime Environment,* 230(2), 444-453.

Wong, M. O., Zheng, Z., Liang, H., Du, J., Zhou, S., & Lee, S. (2023). How does VR differ from renderings in human perception of office layout design? A quantitative evaluation of a full-scale immersive VR. *Journal of Environmental Psychology*, *89*, 102043.

Ye, Y., Wong, S. C., Li, Y. C., & Choi, K. M. (2023). Crossing behaviors of drunk pedestrians unfamiliar with local traffic rules. *Safety Science*, *157*, 105924.

Ye, Y., Wong, S. C., Li, Y. C., & Lau, Y. K. (2020). Risks to pedestrians in traffic systems with unfamiliar driving rules: A virtual reality approach. *Accident Analysis & Prevention*, *142*, 105565.

Ye, Y., Zheng, P., Liang, H., Chen, X., Wong, S. C., & Xu, P. (2024). Safety or efficiency? Estimating crossing motivations of intoxicated pedestrians by leveraging the inverse reinforcement learning. *Travel Behaviour and Society*, *35*, 100760.

Ye, Y., Che, Y., & Liang, H. (2024). Exploring the Influence of Pedestrian Attitude, Propensity, and Risk Perception on Gap Acceptance Between Platooning Autonomous Trucks. *In 2024 IEEE 27th International Conference on Intelligent Transportation Systems (ITSC)* (pp. 3645-3650). IEEE.

Yoon, T., Choi, M., & Lee, S. (2025). Pedestrian perceived risk of construction obstructions and barriers identified via image segmentation. *Applied Sciences*, 15(10), 5261.

Zhu, H., Iryo-Asano, M., Alhajyaseen, W. K., Nakamura, H., & Dias, C. (2021). Interactions between autonomous vehicles and pedestrians at unsignalized mid-block crosswalks considering occlusions by opposing vehicles. *Accident Analysis & Prevention*, 163, 106468.

Zhu, M., Graham, D. J., Zhang, N., Wang, Z., & Sze, N. N. (2025). Influences of weather on pedestrian safety perception at mid-block crossing: A CAVE-based study. *Accident Analysis & Prevention*, 215, 107988.
46

# Appendix

**Appendix A.** Virtual reality sickness questionnaire (VRSQ)

| VRSQ symptom | Oculomotor | Disorientation |
|---|---|---|
| 1. General discomfort | O | |
| 2. Fatigue | O | |
| 3. Eyestrain | O | |
| 4. Difficulty focusing | O | |
| 5. Headache | | O |
| 6. Fullness of head | | O |
| 7. Blurred vision | | O |
| 8. Dizzy (eyes closed) | | O |
| 9. Vertigo | | O |

**Appendix B.** Variables and items

| Variables | Items |
|---|---|
| Violation | I cross the street even though the pedestrian light is red. |
| | I cross diagonally to save time. |
| | I cross outside the pedestrian crossing even if there is one (crosswalk) less than 50 meters away. |
| | I take passageways forbidden to pedestrians to save time. |

| | | |
|---|---|---|
| Error | I cross between vehicles stopped on the roadway in traffic jams. | |
| | I cross even if vehicles are coming because I think they will stop for me. | |
| | I walk on cycling paths when I could walk on the sidewalk. | |
| | I run across the street without looking because I am in a hurry. | |
| Lapse | I realize that I have crossed several streets and intersections without paying attention to traffic. | |
| | I forget to look before crossing because I am thinking about something else. | |
| | I cross without looking because I am talking with someone. | |
| | I forget to look before crossing because I want to join someone on the sidewalk on the other side. | |
| Aggressive behaviors | I get angry with another road user (pedestrian, driver, cyclist, etc.), and I yell at him. | |
| | I cross very slowly to annoy a driver. | |
| | I get angry with another road user (pedestrian, driver, cyclist, etc.), and I make a hand gesture. | |
| | I got angry with a driver and hit his vehicle. | |
| Positive behaviors | I thank the driver who stopped to let me cross. | |
| | When I am accompanied by other pedestrians, I walk in a single file on narrow sidewalks so as not to bother the pedestrians I meet. | |
| | I walk on the right-hand side of the sidewalk so as not to bother the pedestrians I meet. | |
| | I let a car go by, even if I have the right-of-way, if there is no other vehicle behind it. | |
| Risk perception | How dangerous do you think crossing the road in the scenario is? | |
| | What do you think is the probability of having an accident while crossing the road? | |
| | If you were to have an accident while crossing the road in this situation, how severe do you think the consequences of the accident would be? | |
| Concerns about autonomous driving | Do you think that the autonomous truck/car in the video might suddenly lose control and pose a danger? | |

**Appendix C.** Multimodal presence scale (MPS)

| Item |
|---|
| Physical Presence |
| The virtual environment seemed real to me. |
| I had a sense of acting in the virtual environment, rather than operating something from outside. |
| My experience in the virtual environment seemed consistent with my experiences in the real world. |
| While I was in the virtual environment, I had a sense of "being there". |
| I was completely captivated by the virtual world. |
| Social Presence |
| I felt like I was in the presence of another person in the virtual environment. |
| I felt that the people in the virtual environment were aware of my presence. |
| The people in the virtual environment appeared to be sentient (conscious and alive) to me. |
| During the simulation there were times where the computer interface seemed to disappear, and I felt like I was working directly with another person. |
| I had a sense that I was interacting with other people in the virtual environment, rather than a computer simulation. |
| Self-presence |
| I felt like my virtual embodiment was an extension of my real body within the virtual environment. |
| When something happened to my virtual embodiment, it felt like it was happening to my real body. |
| I felt like my real arm was projected into the virtual environment through my virtual embodiment. |
| I felt like my real hand was inside of the virtual environment. |
| During the simulation, I felt like my virtual embodiment and my real body became one and the same. |